\documentstyle[twocolumn,aps,prd]{revtex}
\tighten

\def\bfl{\begin{flushleft}}
\def\efl{\end{flushleft}}
\def\bfr{\begin{flushright}}
\def\efr{\end{flushright}}
\def\bc{\begin{center}}
\def\ec{\end{center}}
\def\be{\begin{equation}}
\def\ee{\end{equation}}
\def\ba{\begin{eqnarray}}
\def\ea{\end{eqnarray}}
\def\baa#1{\begin{array}{#1}}
\def\eaa{\end{array}}
\def\nn{\nonumber }
\def\lb#1{\label{#1}}

\def\drm{d}

\def\lan{{\cal L}}
\def\tlan{\tilde {\cal L}}

\def\Der#1#2{\,\frac{\partial #1}{\partial #2}}
\def\fder#1#2{\,\frac{\delta #1}{\delta #2}}

\def\Fact{\,\verb|!|}

\def\Exp#1#2{\, \text{exp}^{#1}\left[#2 \right] }

\def\ci{\aleph}
\def\od{\wp}
\def\square{{\hbox{$\sqcup$}\llap{\hbox{$\sqcap$}}}}
\def\sqr#1{\square^{#1}}
\def\hfp#1{\,\Delta_F^{(\od)}\!\left( #1\right) }

\begin{document}

\wideabs{
\draft

\title{
High-gradient theories as new essential ingredient for
the beyond-Standard-Model Des(s)ert cuisine:
Getting rid of divergences in Feynman graphs,
unified ``all-in-one'' states and origin of generations
      }

\author{
Konstantin G. Zloshchastiev\cite{z}}

\address{Department of Physics, National University of Singapore,
Singapore 119260, Republic of Singapore
}

\date{~Received: 1 June 2000 ~}
\maketitle

\begin{abstract}
Inspiring ourselves by the assumption that  
the notion of symmetry itself is insufficient
to construct the consistent physics of the Desert 
(the so-called region of energies
beyond the Standard Model)
and some additional insights are needed,
we suggests that high-energy theories must
take into account the higher-order  variations 
of fields.
With this in mind 
we propose the generalization of the concepts of
kinetic energy and free particle.
It is shown that the theory founded on such principles reveals
major features of the genuinely high-energy one,
first of all, it appears to be free from
ultra-violet divergences,
even the self-energy loop terms become finite.
Also we discuss other arising interesting phenomena such as 
the high-gradient
currents and charges, unified ``all-in-one'' multi-mass states
and origin of generations, 
VR symmetry,
regularization-without-renormalization  of SM, etc.
\end{abstract}

}


\narrowtext

Nowadays there is no clear theory describing 
the field-particle habitats of
the region of energies between the electroweak and Planck scales.
Many physicists even name this region
the Desert probably supposing that its nature is
rather meager in comparison with its low-energy (Standard Model) and
high-energy (superstring) borders.
The recently known attempts of generalization of 
SM are concerned with searches for new groups of 
symmetries and carriers of appropriate interactions.
The dream which is inherent to such searches is the one about the
theory which would be genuinely high-energetical in the sense,
e.g., free from ultraviolet divergences \cite{txt-ren}.
For instance, the popularity of supersymmetric theories was arisen 
principally by their
(relatively) good ultraviolet behaviour.
The number of proposed since then symmetries and corresponding models is
so huge that sometimes it is hard to understand whether a paper is
devoted to the group theory or to the high-energy physics.

However, it may happen that 
the notion of symmetry {\it itself} is not enough  
to understand the physics of the Desert 
and some additional, rather
physical than mathematical, insights are needed.
The first insight can appear if we begin to physically
think about the nature of the
Desert.
It is almost doubtless that 
this region is characterized not only by high values of fields
but also by high rates of their variations in space and time.
Therefore, the first assertion we can make is:
(i) 
{\it 
The higher-order field variations 
can not be neglected  in the Desert}.
Further, let us think about the nature
of particles, in particular, about the concept of a free particle
(field).
Indeed, which entity can be regarded as a free (i.e., 
non-interacting) particle in
the region with non-trivial rapidly varying 
polarized vacua and hence with large radiation
corrections and hence with large self-interaction?
How  can we define the system of free fields in the high-energy regime?
Besides, thinking about the mathematical
ways of generalization of
the Standard Model why should they  be based  on the modifying of
interaction (potential)  terms only?
What about the generalization of the kinetic energy itself?
In view of the previous assertion we suggest that
(ii) 
{\it
The usual definition of the free particle becomes 
insufficient in the Desert and needs to be modified},
as well as that
(iii) 
{\it
The notion of the kinetic energy should be generalized}
to take into account the phenomena which appear in the Desert.

So, let us try to materialize these three statements in local Lagrangians.
Let us begin with the generic real scalar field action in $d$-dimensional 
flat spacetime 
\be
S [\phi] = \int \lan (\phi,\partial \phi, ... , \partial^{(\od)} \phi, x) 
\ \drm^d x,
                                                                 \lb{eq01} 
\ee
with the following high-gradient Lagrangian
(throughout the paper we will use the notations of ref.\cite{ryd})  
\be
\lan^{(\od)} = \frac{1}{2} \sum\limits_{k=1}^{\od}
\ci_k 
\left(
      \partial_{\mu_1 \mu_2 ... \mu_k}  \phi
\right)^2 - U,
                                                                 \lb{eq02} 
\ee
where $\od$ is the order of the highest derivative,
$\ci_k$ are some constants of the dimensionality 
$\text{L(ength)}^{2 (k-1)}$ whereas
$[\phi] = \text{L}^{1-d/2}$.
In fact, the model we begin with
belongs to the class of the higher derivative theories which
have a long history \cite{ps} 
but nevertheless below we will try to take a new look at them.
It is evident that this Lagrangian satisfies with
the points (i) and (iii).
To demonstrate the assertion (ii) let us suppose that there are no
(external) interactions that means $U=m^2 \phi^2 / 2$.
Further, assuming for simplicity that  
$
\ci_1 (\partial_{\mu_1} \phi)^2 \ll 
\ci_2 (\partial_{\mu_1 \mu_2} \phi)^2 \ll ... \ll
\ci_\od (\partial_{\mu_1 ... \mu_\od} \phi)^2 
$
we preserve in Eq. (\ref{eq02}) only the highest-derivative term 
to obtain
\be
\tilde\lan^{(\od)}_0 = \frac{1}{2} 
\left(
      \partial_{\mu_1 ... \mu_\od}  \phi
\right)^2 + \frac{(-m^2)^\od}{2} \phi^2,
                                                                 \lb{eq03} 
\ee
where we have fixed $\ci_\od$ and rescaled the scalar field
to be of the dimensionality $\text{L}^{\od-d/2}$.
The equation of motion which follows,
\be
\left[
      \sqr \od  - (- m^2)^\od
\right] \phi = 0,
                                                                 \lb{eq04} 
\ee
contains the usual Klein-Gordon  
because 
$ \sqr \od  - (- m^2)^\od   =
\left[
      \sqr{\od-1}  - (- m^2)^{\od-1}
\right] 
\left(
      \square ~  + m^2
\right),$ etc.
We will call the field/particles governed by such equations
as the
{\it
$\od$-free particles} understanding that the standard
notion of the free particle recovers when $\od \to 1$.
This is the reflection of the suggestion (ii) above.
From the viewpoint of the deterministic principle it means
that the world-line history of a particle becomes depending on
the initial data not only of fields and their first derivatives
but also of their higher derivatives that 
agrees with the assertion (i) and
seems to be true in the highly inhomogeneous and
rapidly varying Desert. 
The satisfaction with the correspondence principle can 
also be seen in the Fourier space 
where Eq. (\ref{eq04}) takes the algebraic form
\be
\left(
      k^{2 \od}  -  m^{2\od}
\right) \phi_k = 0.
                                                                 \lb{eq05} 
\ee
Expanding it in series near the point $\od = 1$ we have
\ba
&&\frac{k^2}{\Lambda^2} 
\left[
      1 + \sum\limits_{n=1}^{\infty}
            \frac{1}{n\Fact }
      \left(
            (\od - 1)
            \ln{\frac{k^2}{\Lambda^2}}
      \right)^n
\right]
-                                            \nn\\
&&\qquad \ \ \ \ \frac{m^2}{\Lambda^2} 
\left[
      1 + \sum\limits_{n=1}^{\infty}
            \frac{1}{n\Fact }
      \left(
            (\od - 1)
            \ln{\frac{m^2}{\Lambda^2}}
      \right)^n
\right] = 0,
                                                                 \lb{eq06} 
\ea
where $\Lambda$ is some characteristic momentum.
One can see that these series converge to the Fourier-image of
the usual Klein-Gordon not only when $\od \to 1$ but
also when $k^2$ and $ m^2 \to \Lambda^2$.
Besides, to obtain, e.g., the exponent $k^{2\od}$ it should be
$k > \Lambda$
otherwise we would have either $\od < 1$ or 
alternating series which converge to
an oscillating function.
Thus, we see that the usual ($\od=1$)-field theories and models
appear to be the low-energy limits of the high-gradient ones
with respect to the characteristic value.

It is interesting that looking at the Fourier expansions above
one can observe that the high-gradient
theories also appear to be the generalization of some theories which do not
have any clear local representation in coordinate space. 
These are, for example, fractional-derivative theories (if we take 
noninteger $2\od$) and logarithmic-momenta theories (if we take
a finite number of terms in the series).
In coordinate space the formers have doubtful (non-local) definition
through integrals whereas the latters have no local derivative 
formulation at all
because appear when considering
dynamical systems with  constraints in the phase space.

Further, 
at even $\od$ there arises the symmetry between virtual tachyons
and real
particles.
The formers are known to do not satisfy with causal on-mass-shell
requirement and have states with complex mass.
The number of the real-virtual partners (including 
``antiparticles'' $m\to -m$) is determined in fact by 
zeros of left-hand side
of Eq. (\ref{eq05}): 
at $\od = 2$ we have VR fourplet $\pm m$, $\pm i m$,
at $\od = 4$ we have VR octuplet 
$\pm m,\ \pm i m, \ \pm \sqrt{i} \, m, \ \pm i\sqrt{i} \, m$, etc.
It is unclear whether VR symmetry is just broken in low-energy regime
(like supersymmetry) or does not exist in Nature at all.
It seems that the latter is more probable because, e.g., 
the $\od$-generalized
propagator for boson fields has non-removable singularities at even 
$\od$, as will be shown below.
Be that as it may, fermionic fields do not have the VR
symmetry: in the Fermi case we have, to a highest-order gradient,
\be
 \tlan^{(\od)} = i 
\bar \psi \gamma^\mu \partial_\mu \sqr{\frac{\od-1}{2}}
\psi - U (\bar\psi,\ \psi),     
                                                                 \lb{eq07} 
\ee
and in the case of ($\od$-)free field, 
$U_0 = (-1)^{(\od+1)/2} m^\od \, \bar \psi\psi$,
we obtain the following $\od$-Dirac equation
\be
\left[
i \gamma^\mu \partial_\mu \sqr{\frac{\od-1}{2}}
 - (-1)^{\frac{\od-1}{2}} m^\od
\right]\psi=0,
\ee
which is meaningful at odd $\od$ only.

{\it Noether theorem}.
The important step we proceed now is the formulation of the notion
of conserved current 
which is crucial both for creating the high-gradient models
with specific symmetries and for functional-integral
quantization of high-gradient theories \cite{cd}.
We demonstrate the Noether theorem for the high-gradient 
scalar field but
the generalization for fields with more spacetime or internal
indices is trivial.
So, let us suppose that the action (\ref{eq01}) is
invariant under the group of transformations
$
{x'}^\mu = x^\mu + \delta x^\mu,$
$
\phi'(x) = \phi(x) + \delta \phi(x), 
$
where the variations
are characterized by the infinitesimal parameter(s) $\delta \omega^\mu$:
$
\delta x^\mu = X^\mu_\nu \delta \omega^\nu, \
\delta \phi(x) = \Phi_\nu \delta \omega^\nu - 
(\partial_\nu \phi)\delta x^\nu, 
$
and we are working in the standard approximation to the order
$O(\delta x^2)$.
Let us give the final result: provided
the generalized equations of motion,
\[
0=\fder{S}{\varphi} \sim
\Der{\lan}{\phi} +
\sum\limits_{k=1}^{\od} (-1)^{k}
\partial_{\mu_1 ... \mu_{k}}
      \Der{\lan}
          {(\partial_{\mu_1 ... \mu_{k}} \phi)},
\]
are hold,
the conserved Noether $\od$-current, 
$\partial_\mu J^\mu_\nu = 0$, has the 
following form
\ba
&&J^\mu_\nu = 
\Der{\lan}
    {(\partial_\mu \phi)} 
\Phi_\nu - \theta^\mu_\alpha X^\alpha_\nu +   \nn\\
&&\qquad \  \sum\limits_{k=2}^{\od} 
\check\partial_{\mu_1 ... \mu_{k-1}}
\left[
      \Der{\lan}
          {(\partial_{\mu_1 ... \mu_{k-1} \mu} \phi)} \cdot
      (\Phi_\nu - X^\alpha_\nu \partial_\alpha \phi)
\right],
                                                        \lb{eq09} 
\ea
where
\[
\theta^\mu_\nu = \Der{\lan}{(\partial_\mu \phi)} \partial_\nu \phi
- \delta^\mu_\nu \lan,
\]
and it is assumed that
\ba
&&\check \partial_{\mu_1 .. \mu_j} [ A\cdot B ]
\equiv
(\partial_{\mu_1 .. \mu_{j}} A) \,  B
-
(\partial_{\mu_1 .. \mu_{j-1}} A) \, \partial_{\mu_{j}} B
+ ... \nn\\
&&\qquad\qquad\qquad\quad   
+ (-1)^j
A \, \partial_{\mu_1 .. \mu_{j}} B. \nn
\ea
The conserved $\od$-charge can be defined in a standard way as 
the integral of $J^0_\nu$
over the spatial $(d-1)$-volume $V$.
It should not be forgotten that,
first, the Noether current can obtain (or loose)
extra indices in dependence 
on indices of both fields and transformation generators,
and, second, that 
the symmetrization over its indices can always be imposed.

Further, 
at the pure translations ($X^\mu_\nu = \delta^\mu_\nu,\ \Phi = 0$)
we have the $\od$-generalized energy-momentum tensor
\be
T^\mu_\nu = 
\theta^\mu_\nu +
\sum\limits_{k=2}^{\od} 
\check\partial_{\mu_1 ... \mu_{k-1}} \!
\left[
      \Der{\lan}
          {(\partial_{\mu_1 ... \mu_{k-1} \mu} \phi)} \cdot
      \partial_\nu \phi
\right],
\ee
which replaces the ordinary one in the Desert,
whereas at the purely 
internal global $U(1)$ transformations of a complex componentless
field,
$
\psi \to e^{-i \Lambda} \psi, \
\psi^* \to e^{i \Lambda} \psi^*, 
$
the conserved current is given 
as a superposition of the usual one
and the sum  of
the high-gradient currents:
\ba
&&J^\mu = i
\left[ 
\Der{\lan}{(\partial_\mu \phi)} \phi
-
\Der{\lan}{(\partial_\mu \phi^*)} \phi^*
\right]
+ \sum\limits_{k=2}^{\od} J^\mu_{(k)}, \nn\\
&&J^\mu_{(k)} = i\;
\check\partial_{\mu_1 .. \mu_{k-1}} \!
\biggl[
      \Der{\lan}
          {(\partial_{\mu_1 .. \mu_{k-1} \mu} \psi)}\cdot \psi \nn\\
&&\qquad\qquad\qquad\qquad\qquad      -                                                    
\Der{\lan}
          {(\partial_{\mu_1 .. \mu_{k-1} \mu} \psi^*)}\cdot \psi^*
\biggr],  \nn
\ea
and it is doubtless that the appearance of high-gradient 
currents for any
given symmetry is inevitable in Desert.

{\it Quantization}.
Once we have the notion of the conserved current
we can quantize the theory using Schwinger's currents 
formulation and functional integral methods \cite{sch}.
Let us begin with the $\od$-Klein-Gordon equation in the presence of
a source current
\be
\left[
      \sqr \od  - (- m^2)^\od
\right] \phi_0 = J,
\ee
hence its solution is
\be
\phi_0 (x) = - \int
\hfp{x-y} J(y) \, d^d y
\ee
where $\hfp{x}$ is the $\od$-Feynman propagator obeying the
$\od$-Klein-Gordon equation with the distributional source
\be
\left[
      \sqr \od  - (- m^2)^\od
\right] \hfp{x} = - \delta^{(d)} (x).
                                                              \lb{eqPropEq} 
\ee
The vacuum-to-vacuum transition amplitude is given by
\[
Z_0[J] = 
\Exp{}{
     - \frac{i}{2}
     \int J (x) \hfp{x-y} J(y) \, d^d x d^d y
    }.
\]
If there exists an interaction 
$
\lan^{(\od)} = \lan^{(\od)}_0 + \lan^{(\od)}_{\text{int}},
$
$
U = - (-m^2)^\od \phi^2/2 + U_{\text{int}}
$
then the complete generating functional
is (up to normalizing factor) 
\be
Z [J] = 
\Exp{}{
     i
     \int \lan^{(\od)}_{\text{int}} 
     \left( 
            \frac{1}{i} \frac{\delta}{\delta J}
     \right) \, d^d x 
    }  Z_0 [J],
\ee
which should be expanded in terms of perturbation series,
Feynman diagrams, and all that.
The crucial point here is the $\od$-Feynman propagator.
Resolving Eq. (\ref{eqPropEq}) we 
obtain (omitting the imaginary Breit-Wigner
terms in the denominator)
\be
\hfp x = \frac{(-1)^{\od-1}}{(2\pi)^d}
\int \frac{e^{-i k x}}{k^{2\od} - m^{2\od}} \, d^d k,
                                                           \lb{eqPropExpr} 
\ee
or, alternatively, performing the Wick rotation we have in Euclidean
space:
\[
\hfp x = \frac{i}{(2\pi)^d}
\int \frac{e^{-i k_E x}}{k_E^{2\od} + (-1)^{\od+1} m^{2\od}} \, d^d k_E,
\]
and considering Eq. (\ref{eq06}) it is easy to see that 
the $\od$-Feynman propagator generalizes the usual one for energies
above (and, by the way, perhaps below \cite{txt1}) 
some characteristic scale $\Lambda$.
The fermionic $\od$-Feynman propagator has a similar form
($\od$ must be odd):
\be
\hfp x = \frac{
               (-1)^{\frac{\od-1}
                          {2}
                    }
              }
              {(2\pi)^d}
\int \frac{e^{-i p x}}
          {p\!\!\!/ 
           p^{\od-1} - m^{\od}
          } \, d^d p,
                                                           \lb{eqFPropExpr} 
\ee
following the $\od$-Dirac equation above.

{\it Unified all-in-one mass states and origin of generations.}
Let us consider first the bosonic case.
If instead of the simplified theory (\ref{eq03}) one considers the
more general one (\ref{eq02}) then it removes (completely or particularly) 
the degeneration of the propagator's pole:
\be
{1 \over k^{2\od}-m^{2\od} } \to  {1 \over
         (k^{2} - m_1^2)^{a_1} ...
         (k^{2} - m_N^{2})^{a_{N_\od}}},
\ee
where $a_i \ (i=1,\ 2, . . .,\ N_\od)$ are the multiplicities of roots,
$1 \leq N_\od \leq \od $ becomes the number of multiple-mass states,
$m_i = m_i (\ci_j,m)$ are masses of states.
Thus, in principle one particle degree of freedom 
can have  several $1 \leq N_\od \leq \od $  mass states
which will be therefore named as the  unified ``{\it  all-in-one}'' 
multi-mass states.
It is interesting phenomenon and perhaps it can be the natural
explanation of such mysteries as the origin of masses and the
existence of several generations of the particles (as well as 
oscillations between them) which have
very similar characteristics (charge, spin, etc.) but 
sharply different values of masses.
At least, it explanates why the maximal number of fermionic generations
is an odd number.
The maximally possible  number of generations is determined by that
of the fermionic
$\od$-Feynmann propagator's poles (fully non-degenerate
case), therefore, the more we 
deepen into the Desert the bigger are $\od$'s and hence the number
of generations.
One may object that the fermionic propagator in SM has only one pole
whereas the number of generations is three.
However, it should not be forgotten that SM 
is infinite in ultraviolet domain hence
incomplete \cite{txt-ren},
therefore, it ``describes'' rather than
``explanates'' the existing symmetries 
and generations.
This at least means that SM should be updated in its
highest-energy scale margin to take into account
high-gradient corrections, see the program-minimum below.

However the most striking benefit we have obtained is 
the absolute power over the divergences.
Returning to Eq. (\ref{eqPropExpr}) we
see that $\od$-Feynman propagators converge for $\od > d/2$ (bosons) 
and $\od > d$ (fermions).
In fact, it is the consequence of 
the fact that high-gradient
theories properly take into account high-energy corrections.
It immediately means that all the integrals of the theory, including the
self-energy loops, have good (or, at least, controlled) 
ultra-violet behavior because they are given as the products of
propagators
\[
I \sim \int \hfp{x_1 - y_1} ... \hfp{y_k - y_M}
\, d^d y_1 ... d^d y_M,  
\]
in dependence on the topology of a concrete diagram.
It should be noted that $N$-supersymmetry-based models 
have much less well-defined ultraviolet behaviour
in the most realistic case $N=1$ besides there exists the 
problem of superpartners.
Besides, the self-energy loops $\hfp 0$ become finite but, 
unlike $N=1$
rigid supersymmetry theories, not necessary zero and 
hence can be regarded as the fundamental 
charasteristics of vacuum hence of the Nature.

{\it Symmetry}.
Any concrete model within the frameworks of a given
theory can be constructed founding on a particular group of symmetry. 
Even if high-gradient and ordinary models have the same symmetry
their properties (first of all, currents and charges) 
in the general case $\od \not\to 1$ 
are different due to appearance of the high-gradient currents 
(however, it does not mean
yet that the conservation law of, e.g., electrical charge
breaks in high-gradient theories: simply the definition of a charge should
be updated to include high-gradient corrections).
Therefore, when trying to construct the consistent beyond-SM physics 
it is necessary
to understand which symmetry survives in the Desert.
Generally speaking, there we can encounter the fact that
many habitual gauge or even global and discrete symmetries 
(such as $CPT$) become approximate.
On the other hand, there can appear a number of 
explicit and hidden symmetries which are inherent to 
high-derivative systems \cite{bd}
including the high-derivative supersymmetries \cite{wes}
(with the known simplifications caused by nature of Grassmanian variables)
and Riemann-Weyl-Cartan ones (in the vicinity of the
high-energy border of the Desert
where spacetime cannot be supposed flat or even locally flat).
All these circumstances can sufficiently 
complicate the constructing
of physical models but nobody asserts that trips across deserts are easy.
Besides, it should be remembered that we have the large advantage
of possessing the well-defined finite
perturbation theory, therefore,  high-gradient models are 
calculable hence {\it verifiable}
independently of whether they are ``renormalizable'' or no.

Further, let us demonstrate the two viewpoints concerning ways 
of the heuristic constructing of desert models illustrating 
them on the simple
$d$-dimensional example, quantum electrodynamics
(which, however, should be regarded just as a toy illustration
rather than realistic extrapolation 
because SM contains QED only as a part of the 
underlying non-Abelian theory):
\be
\lan_{QED} = i \bar\psi \gamma^\mu D_\mu \psi
+ m \bar\psi\psi + \lan_{F},
\ee
where 
$D_\mu = \partial_\mu - i e A_\mu$, and
$
\lan_F = (1/4) F_{\mu\nu} F^{\mu\nu} + \lan_{GF},
$
where $\lan_{GF}$ is the gauge-fixing term.
The minimal $\od$-generalization of the theory
leads 
to the redefinition of covariant derivative
such that
\be
\lan_{\od QED} = i \bar\psi \gamma^\mu D^{(\od)}_\mu \psi
+ (-1)^{\frac{\od - 1}{2}} m^{\od} \bar\psi\psi + \ci \lan_F,
\ee
where 
$D^{(\od)}_\mu = \partial_\mu \sqr{\frac{\od-1}{2}} - i e A_\mu$, 
$\ci$ is some constant of the dimensionality 
$\text{L}^{2(\od-1)}$,
and 
$[\psi]=\text{L}^{(\od-d)/2}$,
$[A_\mu]=\text{L}^{2-\od-d/2}$.
One can see that the $D^{(\od)}$-term
preserves the global $U(1)$ symmetry but
explicitly breaks the gauge one.
Then the first viewpoint tells the following:
such a gauge symmetry does survive in the Desert and 
hence additional terms should
be inserted in the Lagrangian to restore it 
that can easily be done.
The previous definitions of  
currents and charges are only approximate, in high-energy
regime they should be replaced
with the $\od$-generalized ones.
Unlike the first point of view the second one asserts that: (a) this
symmetry eventually
dies in the Desert, (b) this Lagrangian satisfies with the
correspondence principle,
therefore, it can be considered ``as is'',
and hence the main questions are where it goes, what are its symmetries,
conserved values, etc.
Of course, the represented opposite views are quite radical,
and it is clear that the correct way should lay somewhere between them.  

{\it Program-minimum}.
All the above-mentioned recipes of the consistent constructing of the 
Desert field theory give in aggregate the program-maximum.
It is necessary also to outline 
the minimal or ``{\it regularization without
renormalization}'' program  
which is applicable in the lower-bound margin
of the Desert and
less painful because it can be applied
to the Standard Model in the present form without the full
$\od$-generalization
of matter (fermionic) or gauge fields.
Let us recall the set of SM particles:
(i) gauge bosons $W^\pm$, $Z$, $A$ which are expressed as  
superposition of $SU(2)$ gauge triplet $W^a$ and $U(1)$ field B,
(ii) fermions: three generations of leptons and quarks,
(iii) eight gluons $G^a_\mu$,
(iv) Higgs doublet,
(v) Faddeev-Popov 
ghosts (unphysical scalar particles with the Fermi-Dirac statistics
which were introduced to preserve unitarity \cite{fp}):  
eight gluon's ones and four ghosts of the gauge bosons.
The ghosts are of special interest now, see also ref. \cite{gkl}.
Due to their unphysical properties they can appear only as
internal lines of diagrams and play the role of auxiliary particles.
Therefore, the program-minimum 
suggests the following:
if input-output particles (external lines of Feynman graphs) have 
energies comparable with the SM scale 
it is enough to $\od$-generalize only the internal-line particles
in such a way that the Standard Model would become finite without 
UV cutoff, see \cite{bbs} and references therein.

The final question we mention now is the one about the concrete 
value of $\od$.
The most plausible way would be to declare $\od$ as an auxiliary
field and to minimize action with respect to it.
Unfortunately, nobody knows how to variate the derivative 
order 
(here we are not considering the tricks such as transition
to the Fourier space where theory becomes defined even 
for non-integer $\od$ \cite{txt1})
so the tentative recipe is to conduct calculations assuming
$\od$ as general as possible,
and then in final expressions to choose the value
either by involving theoretical considerations such as 
minimization of energies,
preserving of fundamental symmetries, etc., or
directly comparing with experimental data.

\def\AnP{Ann. Phys.}
\def\APP{Acta Phys. Polon.}
\def\CJP{Czech. J. Phys.}
\def\CMPh{Commun. Math. Phys.}
\def\CQG {Class. Quantum Grav.}
\def\EPL  {Europhys. Lett.}
\def\IJMP  {Int. J. Mod. Phys.}
\def\JMP{J. Math. Phys.}
\def\JPh{J. Phys.}
\def\FP{Fortschr. Phys.}
\def\GRG {Gen. Relativ. Gravit.}
\def\GC {Gravit. Cosmol.}
\def\LMPh {Lett. Math. Phys.}
\def\MPL  {Mod. Phys. Lett.}
\def\NPh  {Nucl. Phys.}
\def\PhE  {Phys.Essays}
\def\PhL  {Phys. Lett.}
\def\PhR  {Phys. Rev.}
\def\PhRL {Phys. Rev. Lett.}
\def\PhRp {Phys. Rept.}
\def\NCim {Nuovo Cimento}
\def\TMF {Teor. Mat. Fiz.}
\def\prp {report}
\def\Prp {Report}

\def\jn#1#2#3#4#5{{#1}{#2} {\bf #3}, {#4} {(#5)}} 

\def\boo#1#2#3#4#5{{\it #1} ({#2}, {#3}, {#4}){#5}}



\begin{references}
\bibitem{txt-ren}
It 
is indeed a crucial property because the theory which is merely
renormalizible  is always incomplete in the sense  
needs for the parameters which do not follow
from it (and are taken from, e.g.,  experimental data)
because it cannot theoretically explain (predict) their values.
\bibitem{txt1}
We 
are emphasizing on the word ``below'' because, strictly speaking,
recently also 
there is no sure (supported experimentally) theory 
of the very low-energy particle physics
with energies less than $1$ eV,
and one may recall that SM has also 
the problem of infra-red divergences.
In this connection the generalized propagator above
admits the continuation on the whole axis (or even 
complex plane) of $\od$'s, and thus can
work not
only ``beyond'' but also ``before'' SM
even if there is no formalism in terms of
local values in coordinate space.
\end{references}
\end{document}